\documentclass[aps,prl,twocolumn,amsmath,amssymb]{revtex4}
\usepackage{graphicx}
\usepackage{dcolumn}
\usepackage{bm}

\begin{document}

\title{Tuning  the Spin Hall Effect in a Two-Dimensional Electron Gas}

\author{R. Raimondi$^a$ and P. Schwab$^b$}
\affiliation{$^a$CNISM and Dipartimento  di Fisica "E. Amaldi", Universit\`a  Roma Tre, 00146 Roma, Italy\\
$^b$Institut f\"ur Physik, Universit\"at Augsburg, 86135 Augsburg, Germany}

\date{\today}

\begin{abstract}
We provide a theoretical framework for the electric field
control of the electron spin in systems with diffusive electron
motion. The approach is valid in the experimentally important case
where both intrinsic and extrinsic spin-orbit interaction 
in a two-dimensional electron gas are present simultaneously. 
Surprisingly, even when the extrinsic mechanism is the dominant driving
force for spin Hall currents, the amplitude of the spin Hall
conductivity may be considerably tuned by varying
the intrinsic spin-orbit coupling via a gate voltage.
Furthermore we provide an explanation of the
experimentally observed out-of-plane spin polarization in a (110) GaAs
quantum well.
\end{abstract}

\maketitle

Spintronics aims at exploiting the spin degree of freedom of the
electron to manipulate, store and transfer information \cite{awschalom2007,fert2008}. 
The spin Hall effect \cite{dyakonov1971,dyakonov1971a,hirsch1999,zhang2000,murakami2003,sinova2004}, 
in this respect, has raised great expectations
since it
allows the control of the electron spin by purely electrical means via the spin-orbit interaction.
When the electric field responsible for the spin-orbit interaction is due to
impurities, one often refers to extrinsic effects,
whereas intrinsic ones are associated to electric fields 
due to the band or device structure. 
While the theory of the spin Hall effect mainly concentrated on the separate action of the
two mechanisms so far
\cite{inoue2004,mishchenko2004,raimondi2005,khaetskii2006,malshukov2005,engel2005,tse2006},
in experiments intrinsic and extrinsic spin-orbit coupling are always
present simultaneously.
On the other hand, the first theoretical studies of the interplay
of different mechanisms gave
conflicting results \cite{tse2006a,hu2006,hankiewicz2008}. 
In this paper we provide a theory based on diffusion and drift equations. 
We take the expressions for the spin current and the associated
continuity equation as a starting point for our analysis 
and postpone the sketch of the microscopic derivation to the end of the paper.  
We show that the intrinsic spin-orbit coupling may 
tune the amplitude of the spin Hall currents that are generated by
the extrinsic mechanism,  the  tuning parameter being provided by the 
ratio of the spin relaxation 
times due to the
Dyakonov-Perel mechanism and spin non-conserving scattering.
We settle the issue of the non-analytic behavior of the spin Hall conductivity
when the intrinsic spin-orbit coupling strength goes to zero \cite{tse2006a,hankiewicz2008}.  
To demonstrate the power of our formalism we also examine aspects of current-induced
spin polarization.


We consider a \textit{disordered } two-dimensional electron gas (2DEG) with Hamiltonian  
\begin{equation}
\label{hamiltonian}
H=\frac{{\bf p}^2}{2m}-\frac{{\bf A}\cdot {\bf p}}{m}
  + V({\bf x})-\frac{\lambda_0^2}{4}\boldsymbol{\sigma} \times \partial_{\bf x} V({\bf x})\cdot {\bf p}.
\end{equation}
The intrinsic spin-orbit interaction enters in the form of a spin-dependent
[$SU(2)$] vector potential. For example for the Rashba model the
$SU(2)$ vector potential is given by 
\begin{eqnarray}
  {\bf A}  & = & \alpha m \ {\boldsymbol \sigma}\times { \bf \hat e}_z
        \equiv \frac{1}{2}{\bf A}^a\sigma^a, \ a=x,y,z. \label{vectorpotential}
\end{eqnarray}
$V({\bf x})$ is the scalar potential due to the scattering from
the impurities and gives rise to the extrinsic spin-orbit interaction with the
strength  being characterized by the length $\lambda_0$.
The diffusive limit implies that
the disorder broadening of the energy levels $\hbar/\tau$ must be small compared to
the Fermi energy, but large compared to the spin-orbit energy ${\bf A}
\cdot {\bf p }_F /m$. For simplicity, in the following,  we choose units such that $\hbar =1$ and $c=1$.

In the diffusive limit the spin current polarized in $a$-direction and
flowing along the $i$-direction takes the form
\begin{equation}
\label{spincurrent}
 {j}^{a}_i = \mu s^a {E}_i -D {\tilde \partial}_i  s^a +
 \sigma^{sH}_{\rm ext} \varepsilon_{iab} E_b + \gamma_{ij}^a  E_j
;\end{equation}
the first three terms on the RHS combine the
Dyakonov-Perel \cite{dyakonov1971} expression for the spin current in isotropic systems
with the Kalevich-Korenev-Merkulov \cite{kalevich1994} theory for systems with
linear-in-momentum spin orbit coupling as discussed in
Ref. \cite{korenev2006}; more general spin-orbit fields have been
investigated, e.g., in \cite{malshukov2005}. In our model the spin mobility  and 
the diffusion constant are 
$\mu = -e \tau/m$ and  $D=
\frac{1}{2} v_F^2 \tau\equiv \frac{1}{2} v_F l $, respectively. The spatial derivative
appears in a $SU(2)$-covariant form \cite{jin2006,tokatly2008}
\begin{equation}
\label{covariantspace}
{\tilde \partial}_{\bf x} s^a ={ \partial}_{\bf x} s^a +\varepsilon_{abc}{\bf A}^b s^c,
\end{equation}
and $\sigma^{sH}_{\rm ext}$ is the spin Hall conductivity due to the
extrinsic mechanism. Technically,  the extrinsic spin Hall conductity
is the sum of skew scattering and side-jump contributions
$\sigma^{sH}_{\rm ext} = \sigma^{sH}_{SS} + \sigma^{sH}_{SJ}$ with
\cite{tse2006,engel2005}
\begin{equation}
\label{extrinsic}
 \sigma^{sH}_{SJ}  =  \sigma_D \frac{\lambda_0^2}{4}\frac{m}{e\tau},
 \hskip 0.3cm \sigma^{sH}_{SS}  = \frac{1}{4}  (p_F l) ( 2\pi N_0 v_0 ) \sigma^{sH}_{SJ},
\end{equation}
where $\sigma_D = 2 e^2 D N_0 $ is the Drude conductivity,  $N_0$ the
density of states at the Fermi energy and $v_0$ the impurity
scattering amplitude, cf. Eq. (\ref{disordermodel}) below.
Additionally, we have an intrinsic source of spin currents, namely
\begin{equation}
\gamma_{ij}^{a} = - D N_0 \frac{e \tau}{2 m } \varepsilon_{abc} A_i^b A^c_j
;\end{equation}
it is interesting to note that the intrinsic term is proportional to
the $SU(2)$ magnetic field generated by the spin orbit field,
\begin{equation}
\gamma_{ij}^a E_j  = \frac{ \tau \sigma_D}{8me} \varepsilon_{ikj}{B}^a_k
E_j
\, \text{ with}  \, \,   {\bf B}^a = \tilde \partial_{\bf x} \times
{\bf A}^a. 
\end{equation}
This shows that there is a full analogy between the conventional Hall
current for the charge current that is generated by a magnetic field
via the Lorentz force and the spin Hall effect which
is generated by the $SU(2)$ magnetic field.

The continuity equation associated to the spin current (\ref{spincurrent}) reads
\begin{equation}
\label{continuity}
{\partial}_t s^a  + \tilde { \partial}_{\bf x}\cdot {\bf j}^a -
\frac{1}{2}
\varepsilon_{abc}{\bf A}^b \cdot
{ \bf j}^c_{SJ} 
     + \frac{1}{\tau_s} s^a   = 0 
.\end{equation}
Notice the term subtracting one half of the side-jump contribution to
the spin current.
This term is related to the fact that one half of
the side-jump current has its origin in an anomalous velocity
contribution, where the velocity operator depends on the impurity
potential, as discussed in Ref. \cite{nozieres1973} and in the final part of this
paper. Finally $1/\tau_s$  is a tensor
that describes anisotropic spin relaxation.
For the 2DEG of Eq. (\ref{hamiltonian}) the scattering from
impurities conserves the
out-of-plane component of the spin, whereas the
in-plane-components relax isotropically with the rate
$1/\tau_s = (3/2 \tau) ( \lambda_0 p_F / 2 \sqrt{2})^4 $.

As a first application of Eqs.~(\ref{spincurrent}) -- (\ref{continuity}),  we study the spin Hall effect
in the Rashba model. 
The spin Hall effect in the simultaneous presence of intrinsic and extrinsic spin-orbit
scattering has been studied in the literature before
\cite{tse2006a,hu2006,hankiewicz2008} with the
surprising result  that the spin Hall conductivity exhibits a
non-analyticity in the Rashba coupling strength, i.e., the strictly
extrinsic spin Hall conductivity ($\alpha =0 $) cannot be
recovered from the $\alpha \to 0$ limit of the combined theory.
This surprise now finds a natural explanation.
To be specific, let us now take the driving electric field along the $x$-direction.
For a uniform system the spin current is found as
\begin{equation} \label{eq10}
j_y^z  =   D (2m\alpha) s^y + \left( \sigma^{sH}_{\rm int } + \sigma^{sH}_{SS} + \sigma^{sH}_{SJ}  \right) E
,\end{equation}
where we introduced the intrinsic spin Hall conductivity
$
\sigma^{sH}_{\rm int} = (e / 8 \pi )( 2 \tau / \tau_{DP} 
)$
with  the Dyakonov-Perel spin relaxation
time given by $\tau_{DP}^{-1} = (2 m \alpha)^2 D $. 
Since for the system we consider
an electric field generates an in-plane spin polarization
\cite{edelstein1990,korenev2006}, Eq.~(\ref{eq10}) is not enough to determine the spin current
and we also need the $s^y$-component of the continuity equation,
\begin{equation}
-  {\rm i } \omega s^y + 2 m \alpha \left(j_y^z - \frac{1}{2}
  \sigma_{SJ}^{sH} E \right)  + \frac{1}{\tau_s} s^y  =   0 
.\end{equation}
Solving now the two equations for zero frequency we find the spin Hall conductivity
($j^z_y = \sigma^{sH} E$)
\begin{equation}\label{eq12}
 \sigma^{sH}  =  \frac{  1/\tau_s}{1/ \tau_s + 1/\tau_{DP}} \left( \sigma^{sH}_{\rm int} +
 \sigma^{sH}_{SS} + \frac{1}{2} \sigma_{SJ}^{sH} \right)
 + \frac{1}{2} \sigma_{SJ}^{sH}
,\end{equation}
and the spin polarization is $s^y= s_E$ with
\begin{equation} \label{eq13}
s_E = - \frac{2m\alpha}{1/\tau_s + 1/\tau_{DP} } 
\left( \sigma^{sH}_{\rm int} + \sigma^{sH}_{SS} + \frac{1}{2}
\sigma^{sH}_{SJ} \right) E 
.\end{equation}
In contrast to what was found in \cite{tse2006a,hankiewicz2008},
our results behave analytically in the limit of vanishing Rashba spin
orbit coupling.
Clearly the ratio between $\tau_s$ and $\tau_{DP}$ is an important
parameter in the theory. The authors of Refs. \cite{tse2006a} and
\cite{hankiewicz2008} strictly restricted themselves to the first order
contributions in $\lambda_0^2$ to the spin Hall conductivity. Since
$1/\tau_s \sim \lambda_0^4$ this implies that $1/\tau_{DP}  \gg 1/\tau_s
$ and therefore 
the limit of vanishing Rasha spin-orbit coupling cannot be recovered
correctly within that approximation. When $1/\tau_{DP}  \gg 1/\tau_s$
we find $\sigma^{sH} = \frac{1}{2} \sigma_{SJ}^{sH}$
which is consistent with \cite{tse2006a}. 
To estimate the various terms in Eq. (\ref{eq12}), we consider
a GaAs 2DEG  with
electron density $n_s= 10^{12}{\rm cm}^{-2}$ and mobility
$\mu=10^3 {\rm cm}^2 {\rm V s}$, so that the electrical conductivity is
$\sigma_D = 1.6 \times 10^{-4} /{\rm \Omega}$.
By using 
$\lambda_0=4.7 \times 10^{-8}{\rm cm}$, we find for the side-jump
contribution to the spin Hall conductivity the value
$e\sigma^{sH}_{SJ}=1.3 \times 10^{-7}/ {\rm \Omega}$. 
In order to estimate the skew scattering and the intrinsic
contribution we assume an attractive impurity potential with $N_0 v_0 = -\frac{1}{2}$ and 
$\alpha = 10^{-12}{\rm e V m}$
from which we find
$e\sigma^{sH}_{SS}=-4.3 \times 10^{-7}/{\rm \Omega}$ and 
$e\sigma^{sH}_{\rm int}=8.2 \times 10^{-9}/{\rm \Omega}$. 
The intrinsic spin Hall conductivity is thus negligibly small compared to the
extrinsic ones. Nevertheless the intrinsic spin-orbit coupling must
not be ignored, since $\tau_s =8.4 \times 10^3 {\rm ps} \gg \tau_{DP}=
90 {\rm ps}$. 
Particularly interesting is the situation when $\tau_s \approx \tau_{DP}$
which may be achieved in systems with even smaller Rashba constant
$\alpha$ or in the presence of an additional channel for spin
relaxation.
Controlling the value of $\alpha$ with a gate potential allows
then a fine tuning of the spin Hall conductivity, with the possibility of even a
change in sign as a function of the gate voltage.

As a second application we study the spin polarization in the presence of both
a driving electric field and an external magnetic field along the
$x$-direction. The question is whether the in-plane magnetic field
generates an out-of-plane spin polarization. In earlier publications
\cite{engel2007a,milletari2008} this subject was studied for the pure Rashba model
without extrinsic effects with the conclusion that an out-of-plane
spin polarization requires either an angle-dependent impurity
scattering or a non-parabolic energy spectrum.
Here we demonstrate that the out-of-plane polarization is a generic
feature which does not require any special angle dependence of
scattering. Furthermore, we find that the extrinsic spin-orbit coupling
considerably modifies the effect. 
After adding to the Hamiltonian
(\ref{hamiltonian}) a magnetic field term $-\frac{1}{2} {\bf b } \cdot
\boldsymbol \sigma $ with $b_x = g \mu_B B$, we have to include in
Eq.~(\ref{continuity}) a spin precession term and we get, in the uniform limit,
the Bloch equation for the spin density in the form
%
\begin{equation} \label{eq14}
\partial_t {\bf s} = - \hat \Gamma ( {\bf s} - {\bf s}_{eq})
   - {\bf b}_{\rm eff} \times {\bf s}
   + {\bf S}_E
,\end{equation}
where the matrix $\check \Gamma$ with
$\hat \Gamma_{ij} =- D \left({\bf A}^i \cdot {\bf A}^j - \delta_{ij}
{\bf A}^k \cdot {\bf A}^k \right) + (1/\tau_s)_{ij} $ 
describes spin relaxation, ${ \bf
b}_{\rm eff}$ is the sum of the external and  drift magnetic field
${\bf b}_{\rm eff} = {\bf b } + {\bf b}_D$, $b_{D,i} = \mu  {\bf A}^i
\cdot {\bf E}$,
and ${\bf S}_E$ is an electric field dependent source term.
For the magnetic field in the $x$-direction, one must 
 take into account that the spins do
not relax towards a state with zero spin polarization but to the equilibrium
value $s^x_{eq}= N_0 b_x/ 2$ and $ s^y_{eq}=s^z_{eq}=0$.
%
To the linear order in the electric field, taken along the  $x$-direction,
Eq.~(\ref{eq14}) becomes then
\begin{eqnarray}
\partial_t s^x & = &-\left(\frac{1}{\tau_{DP}}  + \frac{1}{\tau_{s}} \right) 
                 \left( s^x - \frac{1}{2} N_0 b_x  \right)  \label{blochx} \\
\partial_t s^y & = & -\left(\frac{1}{\tau_{DP}} + \frac{1}{\tau_{s}}
\right)(s^y-s_E) + b_x  s^z  \label{blochy}\\
\partial_t s^z &=&-\frac{2}{\tau_{DP}}s^z- b_x s^y + m\alpha \mu
N_0 b_x E    \label{blochz}
,\end{eqnarray}
with $s_E$ being defined in Eq.~(\ref{eq13}).
The stationary spin polarization has then the out-of-plane component
\begin{equation} \label{eq18}
s^z = b_x \frac{  
  \left( 1/\tau_{DP} + 1/\tau_s \right) (m \alpha \mu  N_0 E - s_E)
  }{b_x^2 + (1/\tau_{DP}+ 1/\tau_s)(2/\tau_{DP})}.
\end{equation}
If the zero-field spin polarization $s_E$ and the spin
mobility $\mu$ are fine-tuned such that $s_E = m \alpha \mu  N_0 E$, then there is no
out-of-plane spin polarization. This is the case in the two-dimensional Rashba model
with spin- and angle-independent disorder scattering, since $s_E =- m \alpha \tau e E / 2 \pi  = - \alpha e
N_0 \tau E $, compare Eq.~(\ref{eq13}), and $\mu = - e \tau/m$.
Generically, however, the out-of-plane spin polarization is nonzero. 

Now we turn our attention to the current-induced
spin polarization in a 2DEG of a (110) GaAs quantum well. 
Such a system has been studied experimentally in \cite{sih2005}, where
it was found that an out-of-plane spin
polarization builds up in response to an electric field even in the
absence of external magnetic fields.  To explain this 
experimental finding, it has been suggested
\cite{sih2005}
that the Dresselhaus spin orbit coupling might be
relevant, since the latter points  out of the plane of
the 2DEG in the case of  a (110) quantum well.
To analyze this situation, we add to  Eq. (\ref{hamiltonian}) 
a linear Dresselhaus term, ${\bf A} = {\bf A}_D$, with
\begin{equation}
{\bf A}_D  = - m \beta \sigma^z {\bf e}_y
.\end{equation}
However, ${\bf A}_D$ alone is not enough to explain the
spin-polarization. 
Indeed one
can easily check that the $SU(2)$ magnetic field vanishes,  ${\bf B}^a = \tilde \partial_{\bf x} \times {\bf A}_D^a
=0$, so that ${\bf A}_D$ neither generates spin polarization nor  a
spin Hall current \cite{hankiewicz2006,helix}. 
%
%
In order to obtain a finite effect, we have to keep both the
Rashba and Dresselhaus terms, from which 
we find two non-zero components of the $SU(2)$ magnetic field,
$B_z^z = 2 ( 2 m \alpha )^2 $ and $B_z^x = - 2 (2m\alpha)(2m\beta)$.
Again, the spin dynamics can be described using the Bloch equation (\ref{eq14}): 
the spin relaxation matrix   is now given by
\begin{equation}
\hat  \Gamma = 4  m^2 D
\left(
\begin{array}{ccc}
\alpha^2 + \beta^2 & 0 &  - \alpha \beta \\
0  & \alpha^2 + \beta^2 &  0 \\
-\alpha \beta   & 0  & 2 \alpha^2 
\end{array}
\right)
+
\frac{1}{\tau_s}
\left(
\begin{array}{ccc}
1 & 0 &  0 \\
0  & 1 &  0 \\
0  & 0  & 0  
\end{array}
\right),
\end{equation}
with the vector ${\bf S}_E$ being
\begin{eqnarray}
{\bf  S }_E & = & 
2 m \alpha (\sigma^{sH}_{SS}+ \frac{1}{2}\sigma^{sH}_{SJ}+ 
\frac{\tau \sigma_D }{ 8 m e } B^z_z   )
\left( \begin{array}{c} E_y \\
                       -E_x \\
                        0
\end{array}
\right) \cr
&& + \frac{\tau \sigma_D}{ 4e } B^x_z
\left( \begin{array}{c}
  0 \\
 \beta E_x \\
 -\alpha E_y
\end{array}
\right).
\end{eqnarray}
Due to the absence of the external magnetic field the equilibrium spin
polarization is zero, ${\bf s}_{eq}=0$,
and since we restrict ourselves to the linear response in the electric field
also the drift magnetic field vanishes, i.e.\ ${\bf b}_{\rm eff} = 0$.
Notice that the source ${\bf S}_E$ is zero in the absence of the Rashba term, so that, as
we stated before, the Dresselhaus term alone is not sufficient for
voltage-induced spin polarization. 
While for the electric field in the $x$-direction the spin polarization
remains in the plane, 
an electric field in the $y$-direction generates both in-plane and
out-of-plane components, the latter given by
\begin{equation} 
s^z  = E_y  2 m  \beta \frac{\sigma^{sH}_{SS}+ \frac{1}{2} \sigma^{sH}_{SJ}
+  \frac{\tau e  }{4\pi}(\tau_s^{-1}+4m^2D(2 \alpha^2 + \beta^2) ) }{4m^2D( 2 \alpha^2 +  \beta^2) +2\tau_s^{-1}}
.\end{equation}
The out-of-plane polarization strongly depends on the orientation of
the in-plane electric field, similar to what is found experimentally.
For a quantitative comparison we expect that also the cubic
Dresselhaus term -- which has been neglected in the present study --
will be of importance.
Furthermore the out-of-plane polarization seems to remain non-zero even for
$\alpha \to 0$. This happens because not only the source term goes to
zero but also the spin relaxation rate for the $s^z$ component.

We now   sketch  the microscopic derivation of  the
continuity equation (\ref{continuity}), whose
form  depends, of course, on the definition of the spin current. We use
$ {\bf j}^a = \frac{1}{4} \{  \sigma^a, {\bf v}  \} $,
where ${\bf v}$ includes also an anomalous contribution,
 \begin{equation} \label{eq25}
 {\bf v} = {\bf v}_0 + {\bf v}_{av}, \,  
 {\bf v}_0 = \frac{{\bf p} - {\bf A}}{m },  \, 
 {\bf v}_{av} = -\frac{\lambda_0^2}{4}
 \boldsymbol \sigma \times \partial_{\bf x} V
 .\end{equation}
In addition to the terms in Eq.~(\ref{eq25}), there is also, with the terminology of Ref. \cite{nozieres1973},  
a polarization current, which can be safely ignored
for the diffusive system  under consideration.
Starting from Eq.~(\ref{eq25}), it is not difficult
to derive an {\em exact} continuity equation from the Heisenberg equation of
motion. However, what  we need here is an
equation for the {\em disorder averaged} spin density and current. 
To carry out this more demanding task,  we use a Green's function approach, following Ref.~\cite{raimondi2006}.  
After a gradient expansion the equation of motion for the
Green function, $\check{G}(\epsilon, \mathbf{ p},\mathbf{ x}, t )$, in  the Wigner representation,  reads
\begin{equation}
\label{equationofmotion}
{\partial}_t {\check G}+{\tilde \partial}_{\bf x} \cdot
\frac{1}{2} \Big\{ \frac{{\bf p}-{\bf A}}{m}, {\check G} \Big\}
=-{\rm i} \left[ {\check \Sigma }, {\check G}\right],
\end{equation}
where the check symbol denotes matrix structure in both   Keldysh
 and spin spaces and $[  , ]$ and $\{  ,  \}$ indicate  commutator and anticommutator, respectively. 
The covariant derivative includes both the spin-dependent vector
potential, ${\bf A}$,  as well as the external electric field ${\bf E}$
\begin{eqnarray} 
{\tilde \partial}_{\bf x} ( \cdot ) & = & 
    (\partial_{\bf x}-e {\bf E} \partial_\epsilon ) (\cdot ) 
    -{\rm i}[{\bf A},(\cdot) ] 
.\end{eqnarray}
Physical observables are obtained by integrating over
the energy $\epsilon$ and the momentum ${\bf p}$. For instance, the  spin density polarized along the $a$-axis
is given by the lesser component of the Green function
\begin{equation}
\label{observables}
s^a({\bf x}, t)=
-\frac{{\rm i}}{2} 
  \int \frac{d \epsilon}{2 \pi} \int \frac{d^2 p}{(2\pi)^2} 
  \mathrm{Tr}(\sigma^a { G^<}(\epsilon, {\bf p},{\bf x}, t))
. \end{equation} 
The continuity equation
 is then obtained
after integrating and taking the trace of Eq.~(\ref{equationofmotion}).
The LHS of the equation can be identified with the time
derivative of the spin density plus the covariant derivative of the
contribution to the spin current that is associated with the normal
part of the velocity, 
$\partial_t s^a + \tilde \partial_{\bf x} \cdot {\bf j}^a_0$,
compare Eq.~(\ref{eq25}). 
The RHS of the equation contains the self-energy, which  
 depends on the disorder model and approximations chosen.
Within the Born approximation, the self-energy reads
\begin{equation}
\label{born}
{\check \Sigma}({\bf x}_1, {\bf x}_2)= \overline{ U({\bf x}_1) {\check G}({\bf x}_1, {\bf x}_2)U({\bf x}_2)},
\end{equation}
where   
$U({\bf x}) =V({\bf x})-\frac{\lambda_0^2}{4}\boldsymbol{\sigma} \times \nabla V({\bf x})\cdot {\bf p}$ and the bar indicates the disorder average. 
To perform the disorder average 
we used the standard model of uncorrelated impurities  defined by
\begin{eqnarray}
\overline{ V({\bf x}_1)V({\bf x}_2) } & = & n_i v_o^2\delta ({\bf x}_1-{\bf x}_2),\label{disordermodel}
\end{eqnarray} 
where $n_i$ is the impurity concentration and $v_0$ the scattering amplitude. 
The term in the self-energy which is zero order in the spin-orbit interaction gives rise to the scattering time
$\tau^{-1}=2\pi N_0 v_0^2$, but vanishes after integration over energy
and momentum.
The first order terms are relevant for the side-jump contributions and the
second order term generates the spin relaxation time, 
\begin{equation}
- {\rm i } [ \check \Sigma , \check G ] \rightarrow 
 - \partial_{\bf x} \cdot {\bf j}^a_{av}
 - \frac{1}{\tau_s} s^a.
\end{equation}
Finally, 
when explicitly solving the kinetic equation, one finds that the
anomalous velocity contribution to the current is one half of the side-jump effect, ${\bf j}^a_{av} = \frac{1}{2} {\bf j}^a_{SJ}$.
Putting the pieces together, one obtains the continuity equation (\ref{continuity}).

In summary, we presented a theoretical framework to describe the
spin dynamics in disordered semiconductors in the presence of
spin-orbit coupling both in the bandstructure and impurity potentials.
Although in dirty systems the dominant driving mechanism for the spin
Hall effect is due to skew scattering from impurities,
the size of the effect may be tuned by 
varying the strength of the Rashba coupling via a gate voltage.
We made use of the 
the concept of $SU(2)$ symmetry in a diffusive system. 
This also allowed us to analyse the problem of the current-induced
spin polarization in a (110) GaAs quantum well, showing that 
the Dresselhaus type of spin-orbit coupling alone cannot explain the
experimentally observed effects, but a combination of Rashba and
Dresselhaus terms gives results in reasonable agreement with the
experiment.

We thank  M. Dzierzawa, U. Eckern, C. Gorini, J. Rammer and A.
Shelankov for discussions. This work was 
supported by the Deutsche Froschungsgemeinschaft through SFB 484 and
SPP 1285.

\bibliography{paper}

%
\end{document}